\documentclass[twocolumn,amsmath,amssymb]{revtex4}

\usepackage{graphicx}
\usepackage{dcolumn}
\usepackage{bm}
\usepackage{natbib}

\fontfamily{computermodern}

\usepackage{setspace}

\begin{document}

\title{Giant Anharmonic Phonon Scattering in PbTe}
\author{O.~Delaire}
\author{J.~Ma}
\author{K.~Marty}
\author{A.~F.~May}
\author{M.~A.~McGuire} 
\author{M.-H.~Du}
\author{D.~J.~Singh}
\author{A.~Podlesnyak} 
\author{G.~Ehlers}
\author{M.~D.~Lumsden}
\author{B.~C.~Sales}
\affiliation{ Oak Ridge National Laboratory, 1 Bethel Valley Road, Oak Ridge TN\ 37831
}

\date{\today}

\begin{abstract}
Understanding the microscopic processes affecting the bulk thermal conductivity is crucial to develop more efficient thermoelectric materials. PbTe is currently one of the leading thermoelectric materials, largely thanks to its low thermal conductivity. However, the origin of this low thermal conductivity in a simple rocksalt structure has so far been elusive. Using a combination of inelastic neutron scattering measurements and first-principles computations of the phonons, we identify a strong anharmonic coupling between the ferroelectric transverse optic (TO) mode and the longitudinal acoustic (LA) modes in PbTe. This interaction extends over a large portion of reciprocal space, and directly affects the heat-carrying LA phonons. The LA-TO anharmonic coupling is likely to play a central role in explaining the low thermal conductivity of PbTe. The present results provide a microscopic picture of why many good thermoelectric materials are found near a lattice instability of the ferroelectric type. 
\end{abstract}


\maketitle

Thermoelectric materials are of intense interest for energy applications, because they can transform heat, otherwise lost to the environment, to produce potentially useful electricity \citep{Wood, Goldsmid, Snyder}. The thermoelectric figure-of-merit, $zT=S^2 \sigma T /  \kappa$, which determines the maximum efficiency of the conversion process, is determined by the Seebeck coefficient, $S$, the electrical conductivity, $\sigma$, and the thermal conductivity, $\kappa$ \citep{Snyder}. This formula expresses the fact that the electrical energy produced is in the form of a current driven by the thermoelectric voltage, which, under open circuit conditions, is given by $-S \Delta T$, while heat conduction and electrical resistance are parasitic. Thus, efficient thermoelectric materials must have a low thermal conductivity, in order to preserve the temperature gradient exploited to generate a voltage. The thermal conductivity contains electrical ($\kappa_{\rm el}$) and lattice ($\kappa_{\rm lat}$) components. The
electrical part is related to the electrical conductivity and therefore reducing $\kappa_{\rm lat}$ is central to obtaining high performance
thermoelectrics. 
A powerful approach consists in designing materials with defects and micro/nanostructural features that hinder the propagation of phonons (the atomic vibrations that carry most of the heat in semiconductors), while preserving $\sigma$ \citep{Chen, Goldsmid}. However, the difficulty of fabrication of materials with such tailored nanostructures, and their thermodynamic stability at high-temperature, can be a concern for real-world applications. Thus, thermoelectric materials with an intrinsically low $\kappa_{\rm lat}$ in bulk form are of very high interest. 

PbTe is one of the leading thermoelectric materials in the temperature range 400 to $800\,$K \citep{Heremans}, with  a very low $\kappa_{\rm lat}$, as well as a large Seebeck coefficient when appropriately doped, and a good electrical conductivity \citep{Nolas, Wood}. 
However, the origin of the low $\kappa_{\rm lat}$ is not well understood. PbTe crystallizes in the rocksalt structure, a simple high-symmetry structure \textit{a priori} not associated with low thermal conductivity. The bulk $\kappa_{\rm lat}$ in PbTe is indeed surprisingly low, with $\kappa_{\rm lat}=2 \, {\rm W m}^{-1} {\rm K}^{-1}$  at $300\,$K in single crystalline samples \citep{Akhmedova}, and similar values in polycrystalline samples \citep{Nolas-Goldsmid}. Thus, a perfect single crystal of PbTe is about five times more resistant to heat conduction than Si$_{0.7}$Ge$_{0.3}$ ($\kappa = 10 \, {\rm W m}^{-1} {\rm K}^{-1}$), a thermoelectric alloy with strong random mass disorder. 
Generally, $\kappa_{lat}$ is expected to be low in disordered alloys, materials with nanoscale inhomogeneities, or crystals with complex unit cells containing loosely bound atoms ``rattling'' in cages \citep{Snyder}, all features that can lower $\kappa$ by scattering phonons \citep{Nolas-Yang-Goldsmid, Schweika, Christensen, Koza}. Clearly, another mechanism, unknown until now, must be responsible for the low $\kappa$ in rocksalt PbTe. 

Thanks to recent developments in instrumentation, inelastic neutron scattering (INS) instruments, in particular time-of-flight spectrometers at spallation sources, are now able to measure the entire four-dimensional scattering function, $S({\mathbf Q},E)$, which contains the full information about the microscopic dynamics of materials. These measurements provide much more detail than was previously obtainable with usual triple-axis neutron scattering experiments \citep{Cochran, Alperin}. In parallel, powerful \textit{ab initio} computational techniques can be used to calculate the phonon dispersions, and can be compared directly with experiments.

Here, we present a detailed mapping of the four-dimensional $S({\mathbf Q},E)$ of PbTe based on inelastic neutron scattering measurements on single-crystalline PbTe, and results of \textit{ab initio} phonon calculations using density functional theory (DFT). While our measurements are in general agreement with previous reports \citep{Cochran, Alperin, Daughton}, our more detailed investigations also reveal several new key features that directly relate to the origin of the low $\kappa$. In particular, our measurements of the temperature and wavevector dependences of the TO mode reveal previously unreported anomalies. In particular, we observe the signature of a very strong and extended anharmonic LA-TO coupling. The LA-TO coupling leads to an avoided-crossing behavior in the dispersions, as well as an anomalous lowering and damping of the LA phonons. Our neutron scattering data clearly reveal LA+TO$\rightarrow$LO scattering processes, and an anomalous TO branch at $\Gamma$ displaying a ``waterfall'' effect, as well as a strongly temperature-dependent double-peak structure.  

A single crystal of PbTe ($m\simeq25\,$g) was grown by a modified Bridgman technique, and characterized with x-ray and transport measurements. Details are given in \citep{supplement}. Hall measurements on the crystal yielded a temperature independent carrier concentration of $1.7 \times 10^{17}\,$electrons cm${}^{-3}$ from $4\,$K to $300\,$K, and a room temperature mobility of $1540\,$cm$^2/$Vs, attesting to the high quality of the crystal. Inelastic neutron scattering measurements were performed with the time-of-flight Cold Neutron Chopper Spectrometer (CNCS) at the Spallation Neutron Source, and (on the same sample) with the HB3 triple-axis spectrometer at the High Flux Isotope Reactor, both at Oak Ridge National Laboratory. Details are given in \citep{supplement}. In the CNCS measurements, multiple datasets were acquired for different orientations of the crystal,  and were combined in  software to map the four-dimensional scattering function, $S({\bf Q},E)$, as a function of energy transfer $E$ and wavevector transfer ${\bf Q}={\bf q} + {\bf \tau}$, with ${\bf q}$ a phonon wavevector and ${\bf \tau}$ a reciprocal lattice vector. Our data span multiple Brillouin zones, with larger coverages for higher incident neutron energies, $E_i$. Phonon dispersions were computed from density functional theory (DFT) using the linear response approach, using the same methods as in \citep{An}. Using the phonon energies and polarization vectors from the DFT computations, we calculated the dynamical structure factor for the scattering of neutrons by phonons. Details are given in \citep{supplement}.

\begin{figure*}[tbp]
\center
\includegraphics[width=11cm]{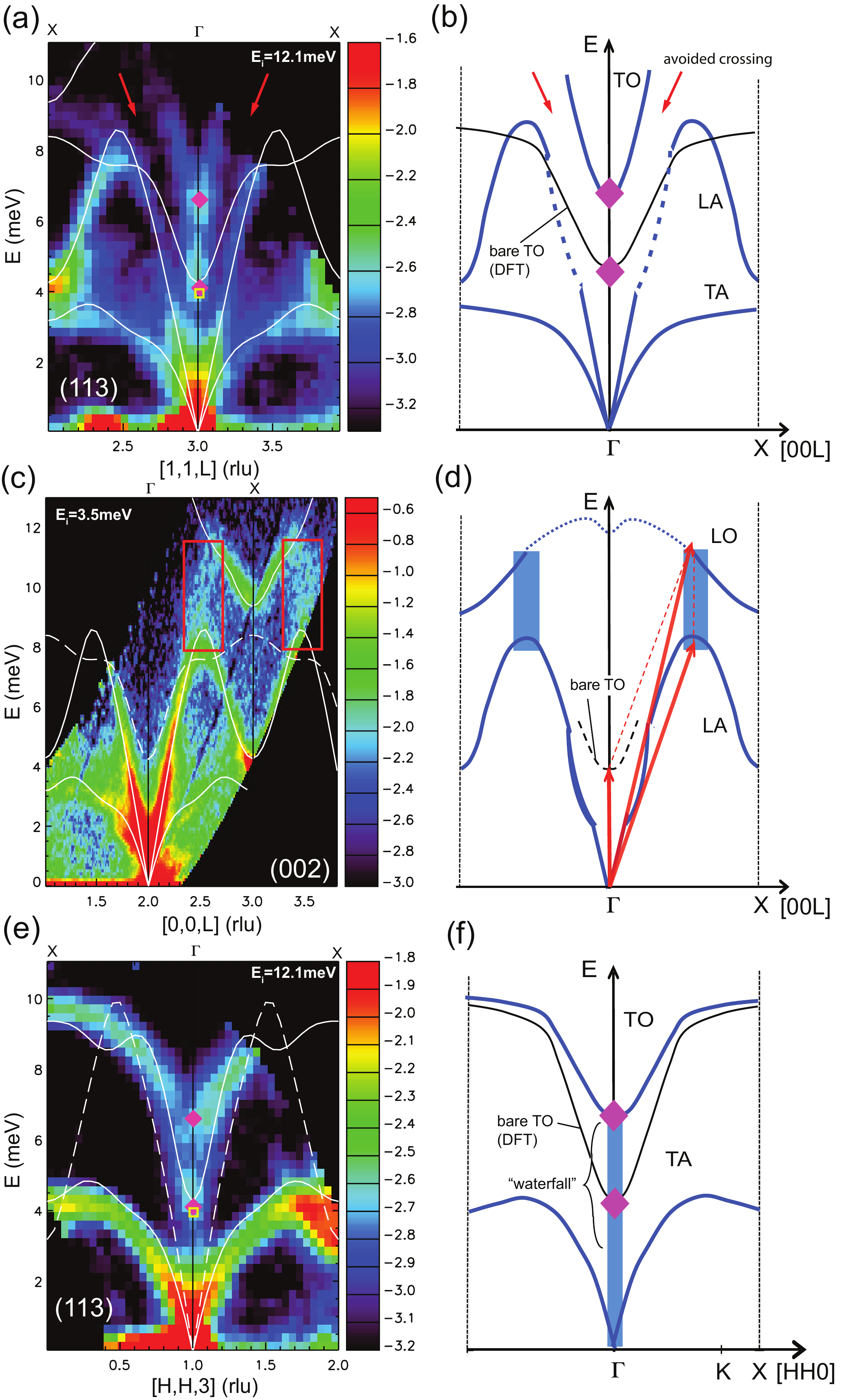}
\caption{CNCS data for PbTe at $300\,$K, showing the avoided-crossing behavior of LA and TO phonon branches in (a), the LA$+$TO$\rightarrow$LO scattering in (c), and the ``waterfall'' effect for the TO branch at $\Gamma$ in (e). Solid and dashed white lines in (a,c,e) are harmonic dispersions calculated with DFT. (a,c): $S(Q,E)$ data for dispersion along [0,0,L] in (113) and (002) zones, respectively. Red arrows in (a) point to the region of the avoided crossing of LA and TO modes. Red boxes in (c) show extra scattering intensity between LA and LO branches. (e): dispersion along [H,H,0] in (113) zone, showing the ``waterfall'' at $\Gamma$.  (b,d,f) are schematics of the dispersions (blue lines), with blue rectangles representing diffuse extra scattering, and the bare TO branch as a thin black line. In all panels, pink diamonds indicate the position of the peaks in the TO scans at $\Gamma$ (113), obtained with HB3. Yellow square in (a,e) is $E_{\rm TO}=3.9 \pm 0.2\,$meV from \citep{Cochran}.  Data in (a,e) were collected with $E_i=12.1\,$meV (phonon creations). Data in (c) were collected with $E_i=3.5\,$meV (phonon annihilations, corrected for detailed balance).} \label{SQE_fig1}
\end{figure*}

One of the striking results of our investigation is the observation of a previously unreported ``avoided crossing'' between LA and TO phonon dispersions, associated with an anharmonic repulsion between these modes. In contrast with the harmonic dispersions calculated with DFT, the measured  LA and TO branches repel each other strongly around ${\bf q}=(0,0,1/3)$. Fig.~\ref{SQE_fig1}-a is for ${\bf q}$ along [00L] in the $(H,K,L)=(113)$ Brillouin zone at 300$\,$K (where (H,K,L) denote reciprocal lattice units, rlu). The white lines plotted on top of the INS data are the harmonic phonon dispersions from DFT. Note that, according to our harmonic DFT calculations, the LA and TO branches should cross at $q=0.4$ (we define $q=L-3$ in this zone) and $E=7.5\,$meV (in agreement with calculations of \citep{Zhang}). Measured LA phonon dispersions are linear for $q\simeq0$, with $E_{\rm LA}$ increasing from $0$ at $\Gamma$ to a maximum $\sim 8\,$meV at $q=0.5$, then curving back down to $\simeq4.5\,$meV at the zone boundary (X), in good agreement with the DFT curves (see data in Fig.~\ref{SQE_fig1}, panels a,c, and corresponding schematics labeling branches in panels b,d). However, the behavior of the TO branch is quite different from the DFT calculations in the harmonic approximation. The transverse optic (TO) branch has a minimum $E_{\rm TO}\simeq 4\,$meV at $\Gamma$, and a steep dispersion to $\simeq10\,$meV at $q=0.2\,$rlu, and does not cross the LA branch. Instead, we observe a strong repulsion between the two branches. The TO branch quickly rises above the top of the LA branch, while the intensity of the LA branch is strongly suppressed around $q=1/3$. We measured the same effect in all other zones where the TO mode was observable ((H,K,L) odd). We note that the LA-TO crossing would normally occur between non-interacting modes of different symmetries along the symmetry lines, and thus the observed avoided crossing must arise from an anharmonic interaction. 

PbTe is an incipient ferroelectric material \citep{Bate, Jantsch}. While paraelectric when undoped, it becomes ferroelectric upon substitution of a few percent of Pb by Sn or Ge \citep{Jantsch}. The ferroelectric transition is of the displacive kind. It arises upon cooling through a critical temperature (Curie temperature, $T_{C}$), at which the ``soft'' TO phonon at the Brillouin zone center ($\Gamma$) slows to zero frequency, freezing in a lower-symmetry ferroelectric structure \citep{Cochran-AdvPhys, Jantsch}. The TO phonon branch in PbTe behaves as in  ferroelectrics, bending to low energies at $\Gamma$, and increasing in energy with increasing $T$ \citep{Cochran, Alperin}. However, the TO mode never becomes fully soft, and the ferroelectric distortion is avoided \citep{Jantsch, Alperin}. Previous first-principles computations of the electronic structure and phonons reported a strong sensitivity of the TO frequency on the volume of the system \citep{An, Rabe, Zhang}, and noted a possible coupling with the longitudinal acoustic (LA) phonons, which are compression waves \citep{An}. Our measurements directly confirm this prediction, and show that the interaction is strong at $300\,$K. 

The LA-TO interaction seen here is reminiscent of the TA-TO interaction in ferroelectric materials, such as perovskite PbTiO$_3$ \citep{Shirane-1970}. However, the TO interaction with LA rather than TA modes is unique to PbTe, to our knowledge, and implies a strong anharmonic mixing, which will not occur at the harmonic level. The repulsion has its origin in an anharmonic coupling term, $V_{\rm anh}$, which renormalizes the harmonic phonon frequencies most strongly at $q$ where bare harmonic dispersions would otherwise cross, and is expected to have the form $\Delta E \propto | \langle {\rm LA} | V_{\rm anh} | {\rm TO} \rangle |^2 / (E_{\rm LA} - E_{\rm TO})$. This mixing also affects the INS intensities, resulting in the LA extinction at $q=1/3$, as we show using mode coupling theory \citep{supplement}.  The effect is observed even more strongly with increasing temperature (see data for $T=600\,$K in \citep{supplement}), with the TO branch raising to higher $E$, and with an additional region of  ``collapsed'' intensity at the bottom of the LA branch, close to X. The increase in the LA-TO interaction with $T$ is expected, since it results from an anharmonic coupling, and the interaction will be stronger as phonon displacement amplitudes increase with $T$.

Figure~\ref{SQE_fig1}-c shows an interaction between LA, LO, and TO branches at the top of the LA branch (in the (002) zone).
The LA dispersion is in good agreement with the DFT results and, in contrast with the (113) zone, its intensity is not extinguished at $q=1/3$ (we define $q=L-2$ in this zone). The absence of LA extinction can be related to the vanishing TO mode structure factor in (002) (see \citep{supplement}). The LO branch is seen clearly for $0.5<q<1$, but is weak for $0<q<0.5$, again in agreement with the dynamical structure factor. At $q=0.5$ (both $L=2.5$ and the symmetry-equivalent $L=3.5$), some extra  intensity is clearly seen between the LA and LO branches (it was also observed in other even zones),  as indicated by red boxes in Fig.~\ref{SQE_fig1}-c. We associate it with a 3-phonon interaction process. The energy difference between LA and LO branches at $q=0.5$ is $4\,$meV, the same as the energy of the TO mode at $\Gamma$. Thus, we can write: LA($q=0.5$) + TO($q=0$) $\leftrightarrow$ LO($q=0.5$), while conserving phonon energy and momentum \citep{Ziman}. The anharmonic LA-TO interaction around $q=1/3$ is also reflected in the LA branch in (002), as a softened and broadened dispersion at low $q$. A more detailed analysis of the LA branch in this region is presented below.

Fig.~\ref{SQE_fig1}-e shows $S({\bf Q},E)$ along [HH0] measured in the (113) Brillouin zone, and reveals a ``waterfall effect'' in the TO branch at $\Gamma$, also arising from the LA-TO interaction. The LA branch is very weak in this zone, but the TA and TO branches are clearly observed, in agreement with calculations \citep{supplement}. The LA becomes visible only close to the X point (1,1,2), where it merges with the TA. While the TA branch in our measurement is in good agreement with previous reports and calculations \citep{Cochran, An, Zhang}, the TO branch is anomalous at the zone center. The branch gradually decreases to $E \simeq 6.5\,$meV with ${\bf q}$ getting close to $\Gamma$, but at $\Gamma$ the dispersion abruptly falls to very low $E$. We could not determine an actual lower bound, but intensity was detected down to $\sim 2.5\,$meV, where it merges with the acoustic branches. This behavior is similar to the ``waterfall'' effect observed in ferroelectric materials, where the strong TA-TO interaction leads to an anomalous TO dispersion near $\Gamma$ \citep{Hlinka, Gehring}. In zones where the acoustic branch has a vanishing intensity, the anomaly is then expected to occur only at $\Gamma$ \citep{Hlinka}, as we observe.  This behavior was seen in both (111) and (113) zones. As a result, the scattering intensity at $\Gamma$ shows a very broad distribution with $E$. We point out that according to our results, the TO branch dips down to low energies as ${\bf Q}  \rightarrow \Gamma$ as generally expected for an incipient ferroelectric, but the LA-TO interaction introduces a complex profile of scattering intensity at $\Gamma$, extending to very low $E$. 

\begin{figure}[tbp]
\center
\includegraphics[width=8cm]{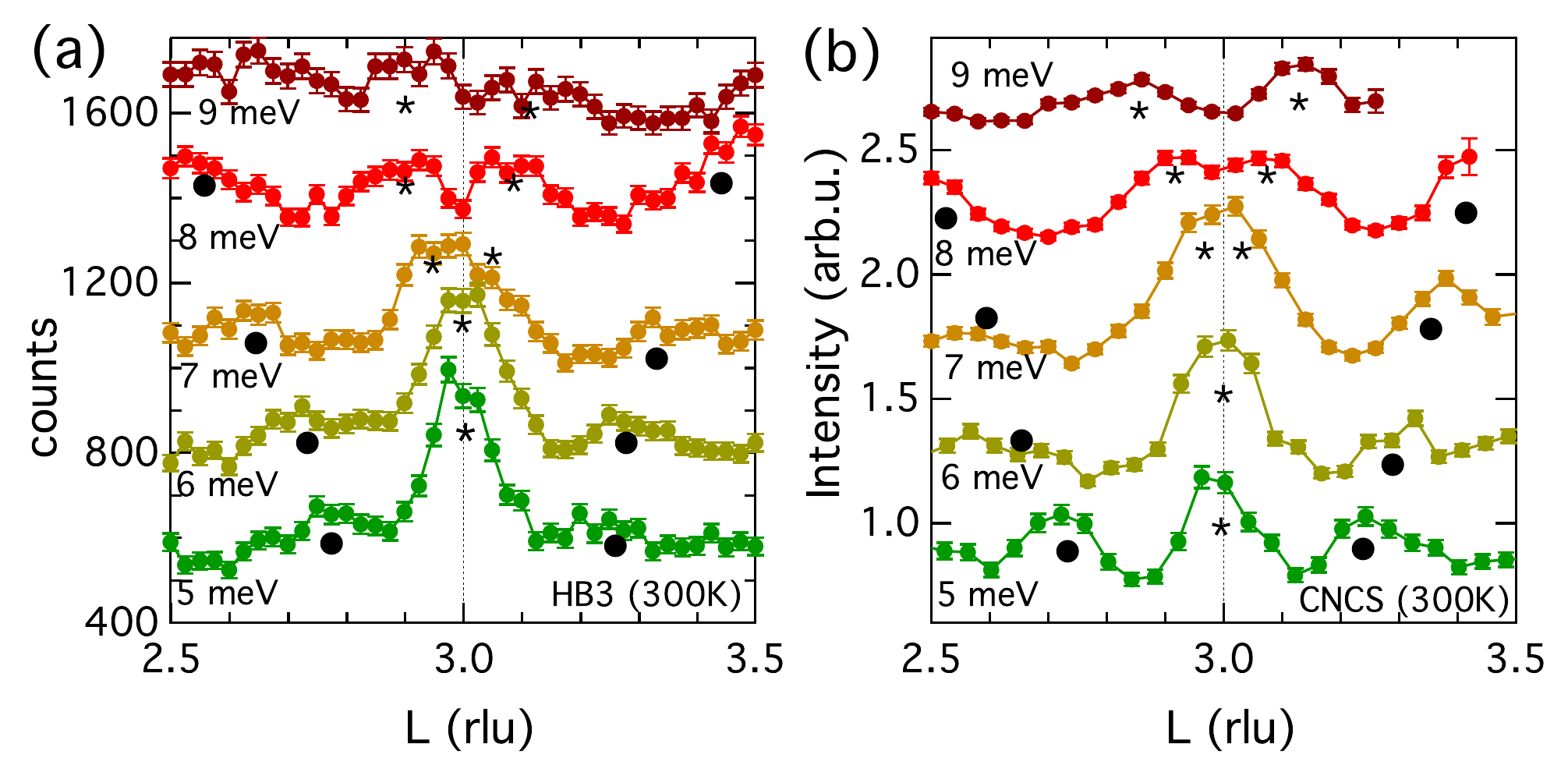}
\caption{Constant-$E$ intensity cuts along [0,0,L] in the (113) zone, showing the avoided crossing behavior of the LA and TO branches. Data were obtained with two different instruments: HB3 (a) and CNCS (b). Both datasets are for the same PbTe crystal at 300$\,$K. Stars (resp. dots) serve to identity the TO (resp. LA) mode positions. The TO mode (stars) is always ``inside'' the LA mode, indicating that the two phonon branches do not cross.} \label{Constant-E_cuts}
\end{figure}

Fig.~\ref{Constant-E_cuts} shows data for constant-$E$ cuts along [0,0,L] in the (113) zone from measurements on both HB3 and CNCS. The two measurements are in excellent agreement, and the avoided crossing of LA and TO branches may clearly be seen. These data confirm our  results shown in Fig.~\ref{SQE_fig1}-a/c. We stress that the anharmonic LA-TO coupling is quite extended in reciprocal space. It is responsible for the ÒwaterfallÓ effect at $q=0$ (in odd zones), the extinction of LA modes and Òavoided crossingÓ with the TO branch around $q=1/3$, and the diffuse scattering intensity at the top of the LA branch ($q=0.5$). This large phase-space for the coupling, together with the strong coupling, is expected to affect the contribution of LA modes to the thermal conductivity of PbTe. Since LA modes are fast-propagating phonons (large group velocities for $q<0.5$), they are expected to be important heat carriers, and thus the LA-TO coupling reported here probably plays an important role in explaining the low $\kappa$ of PbTe.

\begin{figure}[tbp]
\center
\includegraphics[width=8cm]{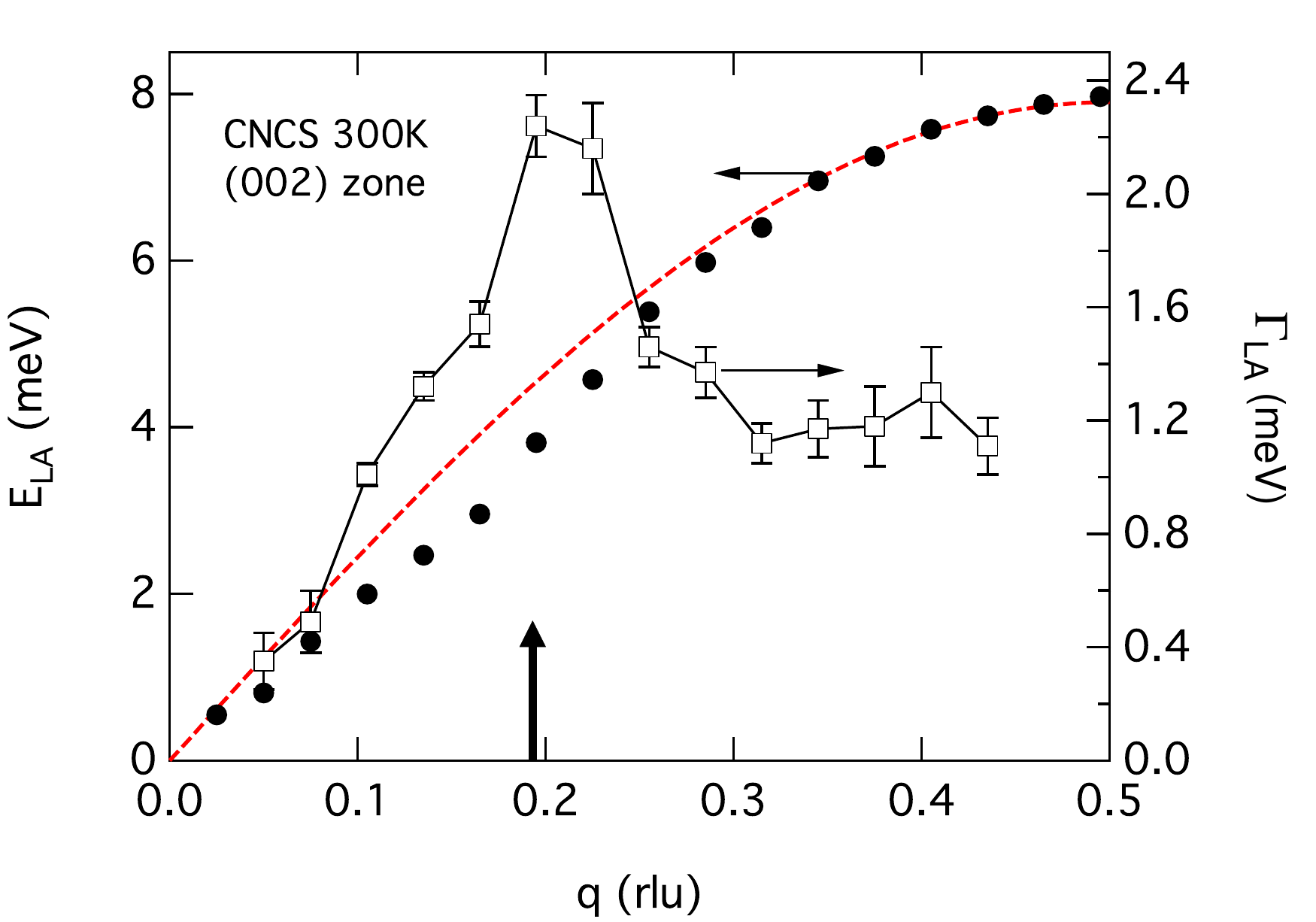}
\caption{LA mode phonon dispersion ($E_{\rm LA}$) and linewidth ($\Gamma_{\rm LA}$, corrected for instrument resolution), showing anomaly at the phonon wavevector $q \sim 0.2$ along [0,0,L]. The data are results of fits of constant-$Q$ cuts through CNCS measurements with $E_i=3.5\,$meV (phonon annihilation processes). The energy  resolution (FWHM) was $0.05\,$meV at $E=0\,$meV, increasing to $0.2\,$meV at $E=8\,$meV. The $q$ resolution was about $0.02\,$rlu. The red dashed line is a guide for the eye ($E=7.9 \times \sin(\pi q)$). Error bars for $E_{\rm LA}$ are comparable to size of markers.} \label{LA_002_dispersion_width}
\end{figure}

We investigated the LA branch in the (002) zone in more details, by performing constant-$Q$ cuts of the CNCS data, and fitting the position of the LA peak (Gaussian fits, corrected for instrument resolution). The results are shown in Fig.~\ref{LA_002_dispersion_width}. The LA dispersion shows an anomalous dip around $q \simeq 0.2$, and the phonon linewidth $\Gamma$ exhibits a peak at the same wavevector, reaching above $2\,$meV at the maximum. Scans measured on HB3 at (0,0,3.8), equivalent to $q=0.2$, also showed a very broad distribution, while clear LA peaks could be observed at $q=0.1$ and $q=0.3$. These anomalies can be directly related to the anharmonic interaction with the TO branch, which results in a down-shift and an increased damping of the LA phonons. We note that in the range $q \simeq 0.15-0.35$, the shape of the LA peak is broad and complex, with extra broad scattering intensity detected between the LA and TA peaks (this was observed in both our CNCS and HB3 measurements). The data for the cuts are shown in \citep{supplement}.

\begin{figure}[tbp]
\center
\includegraphics[width=8cm]{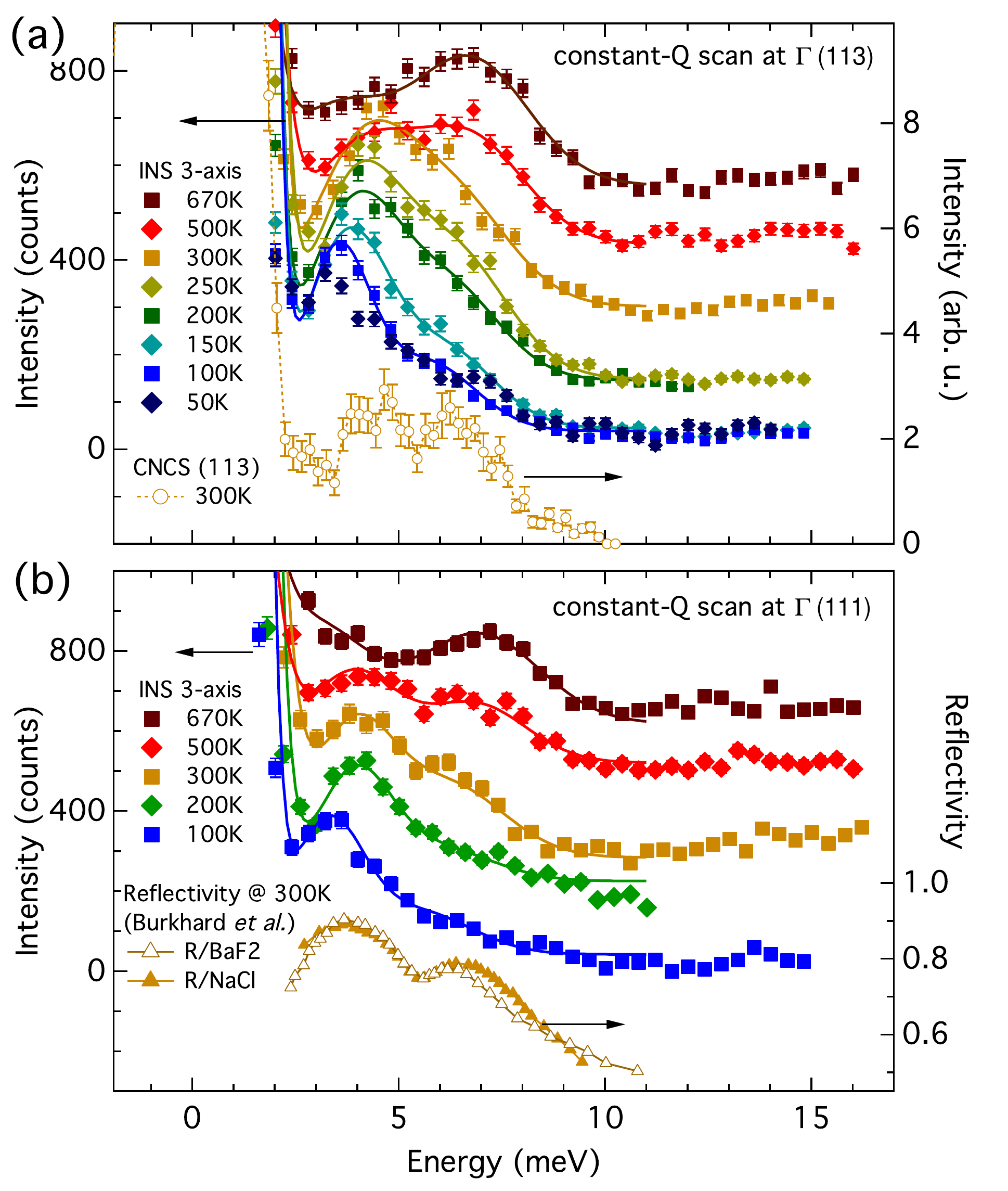}
\caption{Profile of the TO mode in energy at the zone center, measured with HB3 and CNCS, showing the broad double-peak structure and its change with temperature. The neutron scattering intensity (squares and diamonds) {\textit vs} $E$ were measured in constant-{\bf Q} mode at $\Gamma$ points (113) in (a) and (113) in (b), as a function of temperature. Data for different temperatures are offset vertically for presentation. Continuous lines are fits of the data with two Gaussians for the double-peak structure, and an additional Gaussian for the elastic line. Panel (b) also shows the reflectivity data of Burkhard \textit{et al.}, for comparison \citep{Burkhard}.} \label{TO_scans_HB3}
\end{figure}

Investigations of the INS intensity at $\Gamma$ reveal a broad and complex energy profile for the TO mode, as well as a strong dependence on temperature, indicative of anharmonicity. Scans at $\Gamma=$(1,1,1) and (1,1,3) measured with HB3 at multiple temperatures are shown in Fig.~\ref{TO_scans_HB3}. The TO intensity is broad in $E$, in agreement with the CNCS data, and it is also strongly dependent on $T$. At $T=100\,$K, a single well-defined peak is observed at $E=3.2\,$meV, with a broad tail at higher $E$. This peak position agrees well with $E_{\rm TO}$ reported in \citep{Alperin}. With increasing $T$, the peak shifts to higher energy, as occurs in ferroelectrics, and in agreement with previous reports in \citep{Alperin}. Besides this shift, we observe that the INS intensity profile becomes more complex, with a second peak appearing at $E\simeq6\,$meV, while the low-$E$ peak is strongly broadened, corresponding to damping of the TO mode. A double-peak structure is observed at 300K, especially clearly in HB3 data at (111) -- and also in CNCS data at (113)--, and is consistent with the reflectivity data of \citep{Burkhard} (reproduced here for comparison). The high-$E$ peak becomes more intense, and also shifts up in energy with increasing $T$. We fitted the positions of the two peaks with Gaussians, and we find that the low-$E$ peak corresponds to the calculated bare $E_{\rm TO}$ at $\Gamma$, while the high-$E$ peak corresponds roughly to the bottom of the renormalized branch, as seen in Fig.~\ref{SQE_fig1} (pink diamonds). More details are given in \citep{supplement}. We note that since the TO mode is sensitive to strains, different local environments could lead to a broad distribution of TO frequencies. Thus, it is possible that fluctuations (such as reported in \citep{Bozin}), or anharmonic displacements of large amplitudes, could produce the broad TO spectrum.

Our inelastic neutron scattering measurements reveal a strong, unprecedented coupling between LA and TO phonon branches, in particular along the [001] direction. This coupling results in an Òavoided crossingÓ behavior for the measured dispersions around $q \sim1/3$, with a softening and damping of the LA branch and a repulsion of the TO branch. We also observe some intensity in $S({\bf Q},E)$ that can be associated with three phonon process LA$+$TO$\leftrightarrow$LO at $[0,0,q=0.5]$ (involving the TO mode at $\Gamma$), and a ÒwaterfallÓ behavior for the TO branch at $\Gamma (q=0)$. These observations are all consistent with an extended anharmonic interaction between the LA and TO phonons in PbTe, in agreement with DFT calculations \citep{An}. The large phase-space for the coupling, together with the coupling strength, are thus expected to significantly affect the contribution of LA modes to the thermal conductivity of PbTe. The effects described here for PbTe may also be relevant to explain why materials crystallizing in the sodium chloride (rocksalt) structure typically have lower thermal conductivities than materials with cesium chloride structures \citep{Morelli-Slack}. We also point out that the LA and TO modes along [100] involve vibrations of Pb and Te atoms with displacements parallel to [100] directions, and are thus possibly related to the fluctuations recently reported in \citep{Bozin}. 
The TO mode (ferroelectric mode) interacts with acoustic phonons over a wide range of frequencies, and bears some resemblance to a rattling mode, which is a consequence in this case of a strong anharmonic coupling. This provides a mechanism for a material to behave as if it has a very soft lattice for the purpose of heat conduction, while at the same time not being unduly soft from the point of view of stiffness or melting, thus providing an avenue for finding very low lattice thermal conductivity crystalline materials that nonetheless are stable at high temperatures.
The present results provide a microscopic picture for the low thermal conductivity of PbTe and point towards a new direction for finding good thermoelectrics.

\section{Acknowledgements}
We thank M.E. Hagen, J.L. Robertson, and S.E Nagler for helpful discussions. The neutron scattering and theory work was supported by the US DOE, Office of Basic Energy Sciences as part of the S3TEC Energy Frontier Research Center, DOE DE-SC0001299. The Research at Oak Ridge National LaboratoryÕs Spallation Neutron Source and High Flux Isotope Reactor was sponsored by the Scientific User Facilities Division, Office of Basic Energy Sciences, US DOE. B.S. acknowledges funding from DOE Materials Sciences and Technology Division.

\section{Additional Information}

The authors declare no competing financial interests.

\bibliographystyle{is-unsrt}
\bibliography{Delaire_PbTe_arxiv}{}

\end{document}


\title{Giant Anharmonic Phonon Scattering in PbTe - Supplementary Information}
\author{O.~Delaire}
\author{J.~Ma}
\author{K.~Marty}
\author{A.~F.~May}
\author{M.~A.~McGuire} 
\author{M.-H.~Du}
\author{D.~J.~Singh}
\author{A.~Podlesnyak} 
\author{G.~Ehlers}
\author{M.~Lumsden}
\author{B.~C.~Sales}
\affiliation{ Oak Ridge National Laboratory, 1 Bethel Valley Road, Oak Ridge TN\ 37831
}

\date{\today}

\maketitle

\subsection{Sample Preparation and Characterization}

A stoichiometric mixture of high purity Pb (99.999\%) and Te(99.9999\%) shot  (20.72 g Pb, 12.76 g Te) was loaded into a carbonized, round-bottom silica tube, that was then evacuated and sealed. The mixture was heated to $1050^{\circ}\,$C at $2^{\circ}$C/min, homogenized and mechanically mixed for 30 h at temperature, and then cooled to room temperature at $5^{\circ}$C/min. The resulting boule was removed from the silica ampoule and any slight oxidation lightly sanded from the surface.  The boule was broken into mm size pieces, transferred to a silica Bridgman crucible, evacuated and sealed. The Bridgman ampoule was placed in a box furnace with a natural vertical temperature gradient, heated to $1050^{\circ}$C at $2^{\circ}$C/min, soaked for 24 h, cooled over several hours to $980^{\circ}$C, slowly cooled ($0.6-1^{\circ}$C/h) to $750^{\circ}$C, followed by a $1^{\circ}$C/min cool to room temperature. The resulting single crystal was removed from the silica ampoule. Small slices were cut from the crystal using a slow speed diamond saw for additional x-ray, chemical and transport characterization. Hall measurements on the crystals yielded a temperature independent carrier concentration of $1.7 \times 10^{17}$ electrons$/$cm$^3$ from $4\,$K to $300\,$K, and a room temperature mobility of $1540\,$cm$^2/$V s. (The mobility at $4\,$K is an astounding $1.7 \times 10^6\,$ cm$^2$/ V s). 

\subsection{Inelastic Neutron Scattering Measurements}

In the measurements on CNCS, the sample was mounted with an (HHL) plane approximately horizontal. at $T=300, 500, 600\,$K using  a low-background furnace. In CNCS measurements, multiple datasets were acquired for different orientations of the crystal, corresponding to rotations about the vertical ($// [-110]$) axis (typically in steps of 2 or 4$^{\circ}$). The datasets for different rotations were then combined in software to produce a four-dimensional sampling of the scattering function $S({\bf Q},E)$ of the crystal. 
Small departures from the exact horizontal (HHL) orientation were corrected in software when analyzing the data. Several incident energies were used, $E_i=3.51, 12.03, 25\,$meV, with corresponding energy resolutions (full width at half maximum) at the elastic line $2 \sigma_E = 0.05, 0.5 , 1.0\,$meV. At each incident energy and each temperature, multiple datasets were acquired for different orientations of the crystal, corresponding to rotations about the vertical axis (typically in steps of 2 or 4$^{\circ}$). The datasets for different rotations were then combined with the software MSLICE to produce a four-dimensional sampling of the scattering function $S({\bf Q},E)$ of the crystal \cite{Coldea}. These data were then sliced to produce scattering intensity as a function of energy transfer $E$ and momentum $\bf{Q}$ parallel to a chosen direction.

\begin{figure}[tbp]
\center
\includegraphics[width=8.5cm]{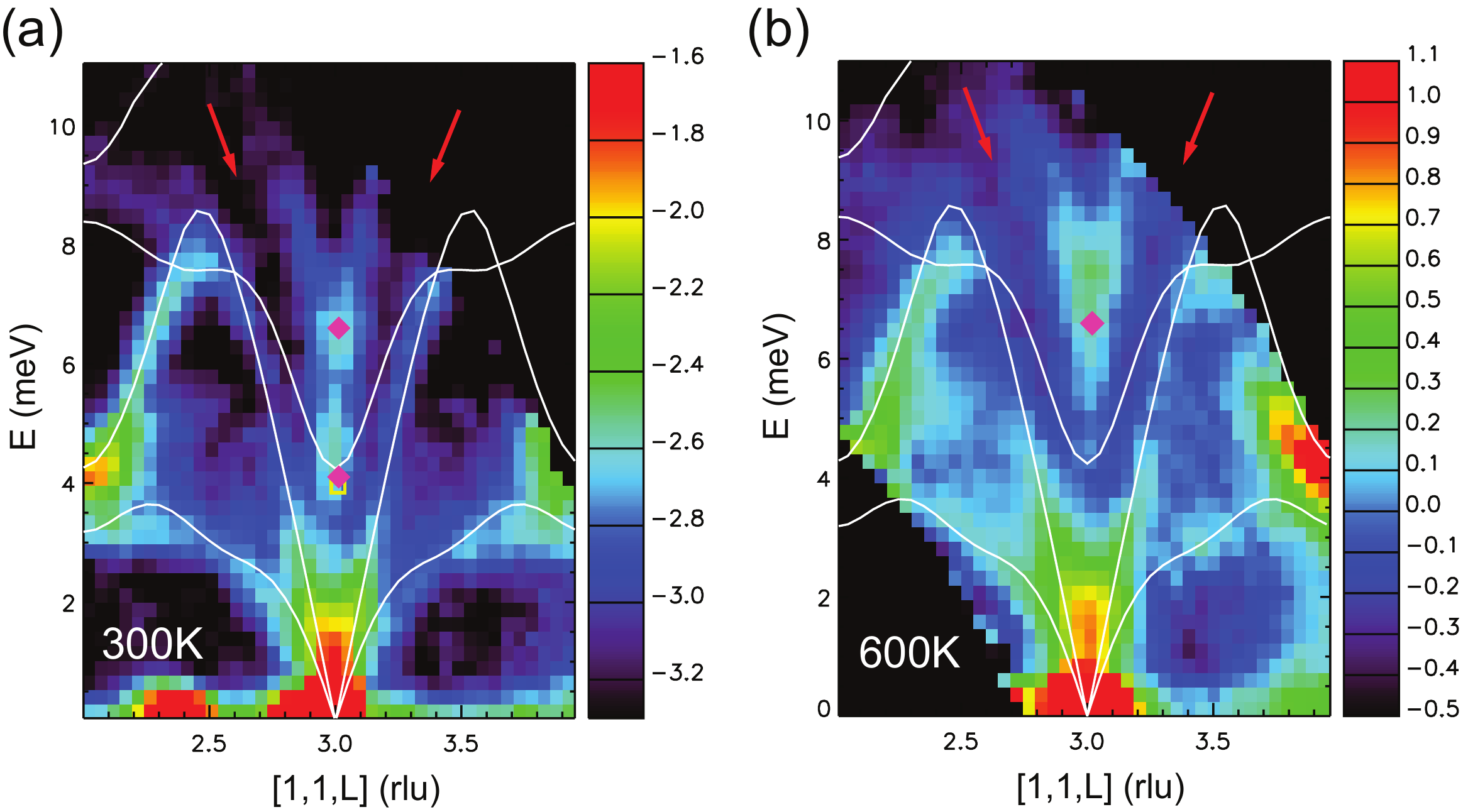}
\caption{CNCS $S(Q,E)$ data for dispersion along [0,0,L] in (113) Brillouin zone, showing the avoided-crossing behavior of LA and TO phonon branches at $300\,$K in (a), and at $600\,$K in (b). White lines are the harmonic phonon dispersions calculated with DFT. The pink diamonds indicate the positions of the peaks in the TO scans at $\Gamma$ (113), obtained with HB3 at $300\,$K and $670\,$K. } \label{SQE_00L_113_vs_T}
\end{figure}

Figure~\ref{SQE_00L_113_vs_T} shows the CNCS $S(Q,E)$ data for dispersion along [0,0,L] in (113) at $300\,$K and $600\,$K. As may be seen on the figure, the avoided crossing of LA and TO branches (see main text) is observed more strongly at $T=600\,$K than at $300\,$K, with the TO branch raising to higher $E$, and with an additional region of  ``collapsed'' intensity at the bottom of the LA branch, close to X. At $600\,$K, the TO branch shows only one maximum of intensity at the $\Gamma$ point, in agreement with the HB3 constant-$Q$ scans.

\begin{figure}[tbp]
\center
\includegraphics[width=8.5cm]{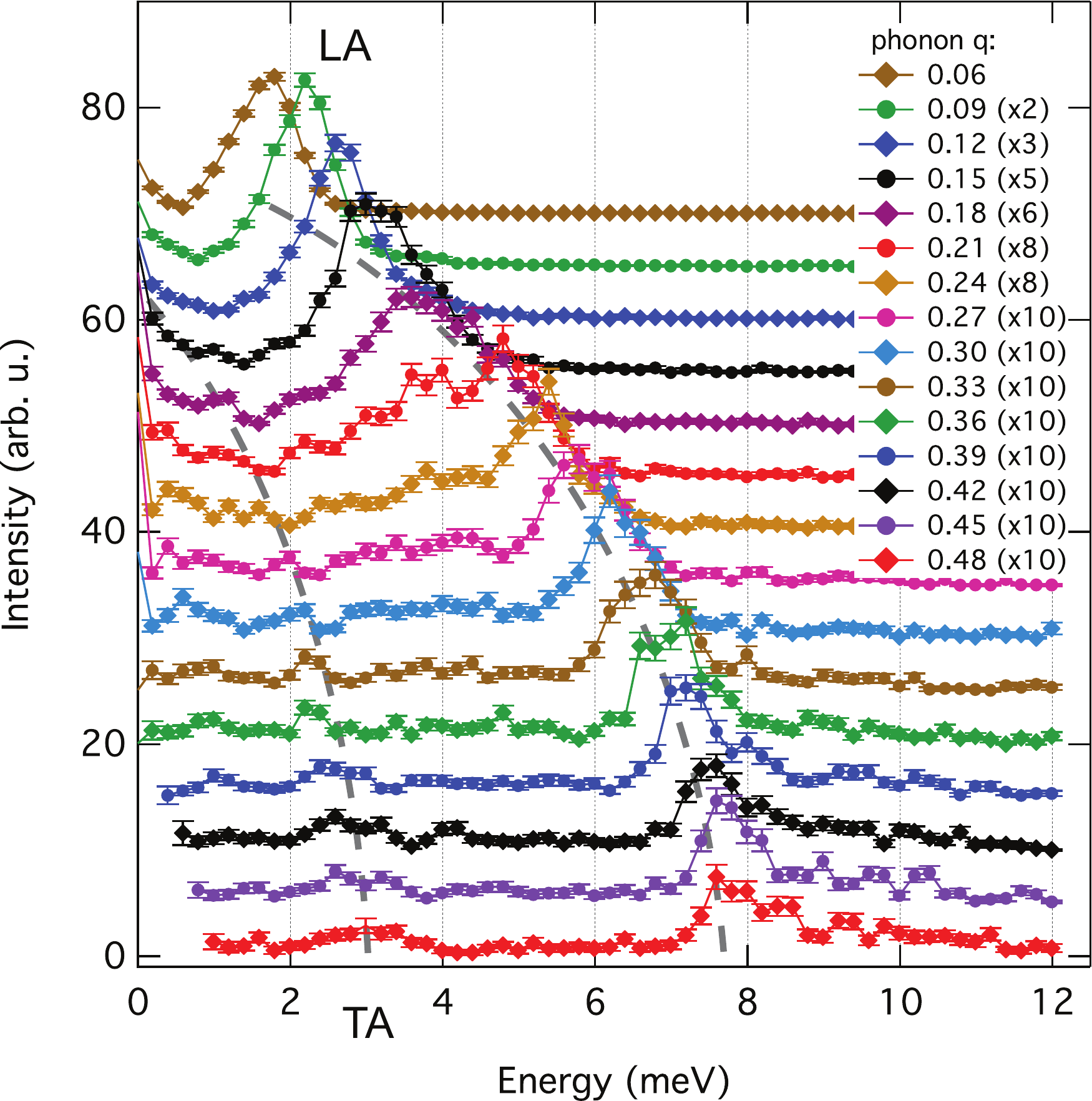}
\caption{(a) Constant-{\bf Q} cuts for $q$ along [00L] in (002) zone from CNCS data at $300\,$K ($E_i=3.5\,$meV). Curves are vertically offset for clarity. Dashed lines are guides for the eye, indicating the positions of the transverse acoustic (TA) and longitudinal acoustic (LA) phonon peaks. The anomaly in the shape and width of the LA peak is clearly seen in the range $q=0.15-0.25\,$rlu.} \label{LA_cuts_CNCS}
\end{figure}

In order to investigate the longitudinal acoustic (LA) phonon dispersion in the (002) zone. We took some constant-$Q$ cuts from the CNCS data in Fig.~1 (main article), integrating of a range of $\Delta Q_{[001]} = 0.03\,$rlu at each step ($\Delta Q_{[110]} =0.15$, and $\Delta Q_{[-110]} =0.15\,$ rlu). Resulting data are shown in Fig.~\ref{LA_cuts_CNCS}. The data in the figure were corrected for detailed balance, and are plotted {\textit vs} positive phonon energy. The LA peak is clearly observed, and one may notice that it is sharp at low-$q$, broadens around $q=0.2$, and then sharpens again to the top of the branch at $q=0.5$. The LA peak position and peak width was fitted with a Gaussian, and the phonon linewidth was obtained as the peak full-width at half maximum, after correction for instrument resolution (including the effect of $Q$-resolution and dispersion slope on the $E$ resolution). Results are plotted in the main manuscript. We note that the linewidths were difficult to extract below $q=0.05$ (because of nearby TA branch) and above $q=0.45$ (because of very broad LA-LO interaction), and are not reported. The transverse acoustic peak (TA) is also observed, although weakly (in agreement with dynamical structure factor calculations, see below), and its position is in agreement with results published in \cite{Cochran}. In the region of the LA-TO avoided crossing ($q=0.1-0.3$), some broad scattering intensity is observed between the LA and TA peaks, which may also be associated with the strong anharmonic behavior in this region.

Additional INS measurements were performed with the triple-axis neutron spectrometer HB-3 at the High-Flux Isotope Reactor at Oak Ridge National Laboratory. We used pyrolitic graphite monochromator and analyzer (in (002) reflection condition), and worked in constant final energy mode, with $E_f=14.7\,$meV. The horizontal collimation angles were 48, 40, 40, and 120 minutes of arc for the Soller collimators prior to the monochromator, between monochromator and sample, between samples and analyzer, and between analyzer and detector, respectively. An (HHL) crystallographic plane of the sample was aligned with the horizontal scattering plane. The sample was mounted either inside an helium cryostat or a furnace for measurements at $T=10, 50, 100, 150, 200, 250, 300\,$K (cryostat), or $T=300, 500, 670\,$K (furnace). The signal from the empty furnace was also measured. A vanadium rod was measured to determine the energy resolution of the instrument at the elastic line (zero energy transfer). The profile was Gaussian, centered on $E=0\,$meV, with a 1.1meV full width at half maximum. 

Scans at $\Gamma=$(1,1,1) and (1,1,3) were measured with HB3 at multiple temperatures (see main article). The HB3 data were fitted with Gaussians for the low- and high-$E$ peaks (and another Gaussian for the elastic peak). The positions of the centers of the two Gaussians are plotted as a function of temperature in Fig. 4. The widths of the excitations were extracted by corrected for the instrument resolution function, which was calculated as a function of energy transfer, and for the instrument conditions used. The peaks observed in $I(E)$ are significantly broader than the instrument resolution (see data in main article). 

\begin{figure}[tbp]
\center
\includegraphics[width=8cm]{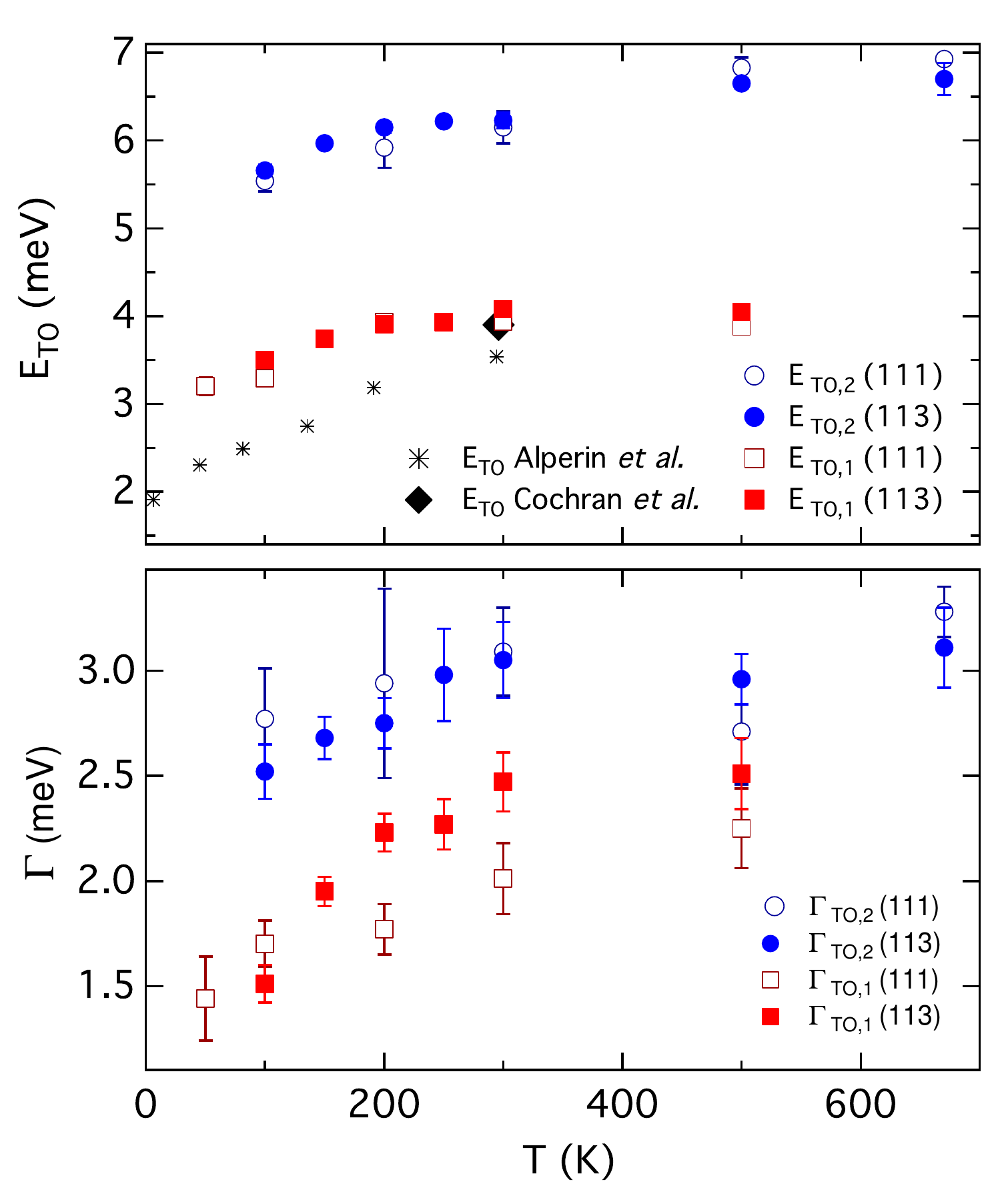}
\caption{Position (a) and linewidth (b) of peaks fitted to HB3 data at the (111) and (113) zone centers.} \label{TO_pos_width}
\end{figure}

We plot our results for the positions of the two peaks as a function of temperature in Fig.~\ref{TO_pos_width}-a. As can be seen on this figure, the position of the low-$E$ peak at $300\,$K coincides with the value of $E_{\rm TO}$ at 296$\,$K previously reported by Cochran \cite{Cochran}. The value reported in \cite{Alperin} for $E_{\rm TO}$ at 295$\,$K is slightly smaller than our measurement (by $0.4\,$meV), but still in good agreement, considering that $E_{\rm TO}$ at the zone center could be sensitive to doping levels. The T-dependence of $E_{\rm TO,1}$ is comparable to that reported in \cite{Alperin}, although we find energies that are consistently slightly higher. The energy of the second TO peak, $E_{\rm TO,2}$, is about 2$\,$meV higher than $E_{\rm TO,1}$, at all temperatures. Both $E_{\rm TO,1}$ and $E_{\rm TO,2}$ follow a behavior reminiscent of that in ferroelectrics, but we could not successfully fit their $T$-dependence to the typical ferroelectric mode behavior ($E^2 \propto (T-T_{\rm C})$) over the full temperature range. Also, we observe an apparent saturation in the behavior of $E_{\rm TO,1}$ and $E_{\rm TO,2}$ at high $T$, which could be related to the saturation in amplitude of the structural flutuations reported in \cite{Bozin}. 

Our results for the widths of the peaks are shown in Fig.~\ref{TO_pos_width}-b. A significant increase of the energy widths of the TO peaks (FWHM with of Gaussian peak fits, corrected for resolution), $\Gamma$, is observed with increasing temperature for both TO peaks. The increase in $\Gamma$ is related to a damping of phonon excitations, and is generally expected for vibrations in anharmonic potentials. As may be seen in Fig.~\ref{TO_pos_width}, the width (FWHM) of the low-energy TO peak ($\Gamma_{\rm TO, 1}$) is more strongly dependent than that of high-energy TO peak ($\Gamma_{\rm TO, 2}$). Between 50 and 500$\,$K, $\Gamma_{\rm TO, 1}$ nearly doubles. At 670$\,$K, we could not extract the position or width of this peak anymore, as it became too broad and the $E_{\rm TO,2}$ became more intense (see main article). The precise $T$-dependence of phonon linewidths in solids is a topic of current interest, and definite trends remain to be established \cite{Fultz}. Both $\Gamma_{\rm TO, 1}$ and $\Gamma_{\rm TO, 2}$ appear to saturate (or at least increase more slowly) as $T$ reaches above 300$\,$K, which is rather unexpected, but could be related to the saturation in the amplitude of fluctuations reported in \cite{Bozin}. It is also interesting to compare the position of the two TO peaks at $\Gamma$ with the dispersion branches reported in \cite{Cochran} and the INS intensity measured with CNCS (main article). The position of the peaks observed at 300$\,$K in HB3 data is shown with pink squares in Fig.~1 (superimposed on the $S(Q,E)$ measured with CNCS). We can see that $E_{\rm TO,2}$ matches the top of the region of diffuse scattering at the zone center, and corresponds to the energy where the dispersion would be expected to converge for $Q \rightarrow 0$. On the other hand, $E_{\rm TO,1}$ matches the position of TO mode in the harmonic DFT calculations.

\subsection{Dynamical Structure Factor and Mode Coupling Calculations}

The phonon frequencies and the polarization vectors were calculated on a fine 40x40x40 grid of $k$-points in reciprocal space using the linear-response approach, following methods in \cite{An}. The lattice parameter used in the calculations was $a=6.376\AA$, which gave  dispersions in close agreement with the measured dispersions at 300K, in particular for the energy of the TO branch close to $\Gamma$ \cite{An, Cochran}. From the calculated phonon frequencies and polarization vectors, we computed the dynamical structure factor in [110] and [001] directions in the (113), (004), and (220) Brillouin zones. The neutron scattering intensity was calculated based on the DFT phonon polarizations $\epsilon_{d,s}$ and energies $E_s$, where $d$ is an index for the atom in the crystal basis (Pb or Te), and $s$ is the phonon mode index, for phonon wavector $q$:
\begin{eqnarray}
S(Q,E) &\propto& \sum_{s} \sum_{\tau} \frac{1}{E_s} \Big| \sum_{d} \frac{\overline{b_d}} {\sqrt{M_d}} \exp{(i {\mathbf Q} \cdot {\mathbf r_d} )} \exp{(- W_d)} \nonumber \\ 
{} & {} &  ({\mathbf Q} \cdot {\mathbf \epsilon_{d,s}} ) \Big| ^2  \times \langle n_s  + 1 \rangle \delta(E-E_s) \delta( {\mathbf Q} - {\mathbf q} - {\mathbf \tau})    \, \nonumber ,
\label{sqe}
\end{eqnarray}
\noindent with ${\mathbf \tau}$ a reciprocal lattice vector, ${\mathbf Q}$ the wavevector transfer, $\mathbf r_d$ the position vector for atom $d$ inside the unit cell, $M_d$ its mass, and $\overline{b_d}$ its coherent cross-section \cite{Squires}. The thermal occupation function is taken into account through the Bose-Einstein distribution $n_s$ (we used $T=300\,$K in our calculations). The Debye-Waller factor $W_d$ was also calculated from the polarizations and energies:
\begin{eqnarray}
W_d & = & \frac{\hbar^2}{4 M_d} \sum_{s} \frac{\big| {\mathbf Q} \cdot {\mathbf \epsilon_{d,s}} \big|^2}{E_s} \langle 2 n_s +1 \rangle
\nonumber
\end{eqnarray}

\begin{figure*}[tbp]
\center
\includegraphics[width=14cm]{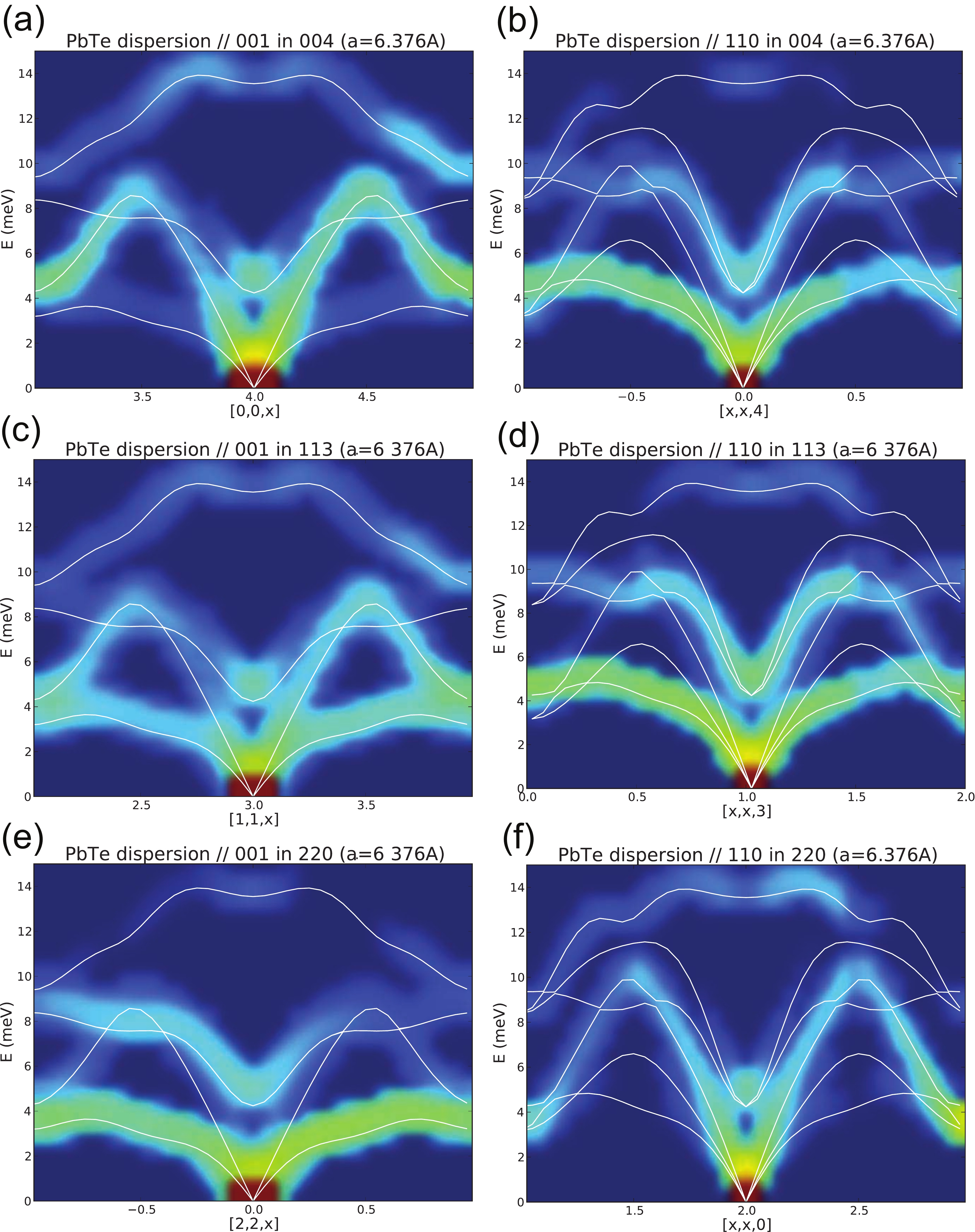}
\caption{Dynamical structure factor (color maps) of PbTe for harmonic phonons, calculated from phonon polarization vectors obtained by linear-response DFT (see text). White curves are harmonic phonon dispersions from linear-response DFT calculations. a: dispersion along [001] in (004) zone. b: dispersion along [110] in (004) zone. c: dispersion along [001] in (113) zone. d: dispersion along [110] in (113) zone. e: dispersion along [001] in (220) zone. f: dispersion along [110] in (220) zone.} \label{StructureFactor}
\end{figure*}

Results are shown in Fig.~\ref{StructureFactor}. The calculated dynamical structure factor is in good agreement with the measurements along [001] in the (002)-(004) zones, with only the LA and LO branches having intensity, and the intensity of the LO becoming stronger as it gets closer to the zone boundary. The main difference is the extra scattering intensity observed experimentally halfway between $\Gamma$ and X between LA and LO,  as discussed in the main text. Similarly, along [110] in the (113) zone, only the TA and TO branches have strong intensity, as observed experimentally, although the  the harmonic calculations do not capture the waterfall effect in the TO  branch at $\Gamma$. The structure factor calculation for harmonic, non-interacting phonons show the strongest discrepancy with measurements for the [001] dispersion in the (113) zone, as discussed in the main text. Although the TO, LA, and TA are all visible, the experiment shows a gap between the LA and TO (avoided crossing) around $q=1/3$, where harmonic calculations predict overlapping intensity. This discrepancy is associated with the LA-TO interaction, and we perform a simple model calculation below to show how this interaction can lead to the extinction of the LA branch near the avoided crossing.

\begin{figure*}[tbp]
\center
\includegraphics[width=14cm]{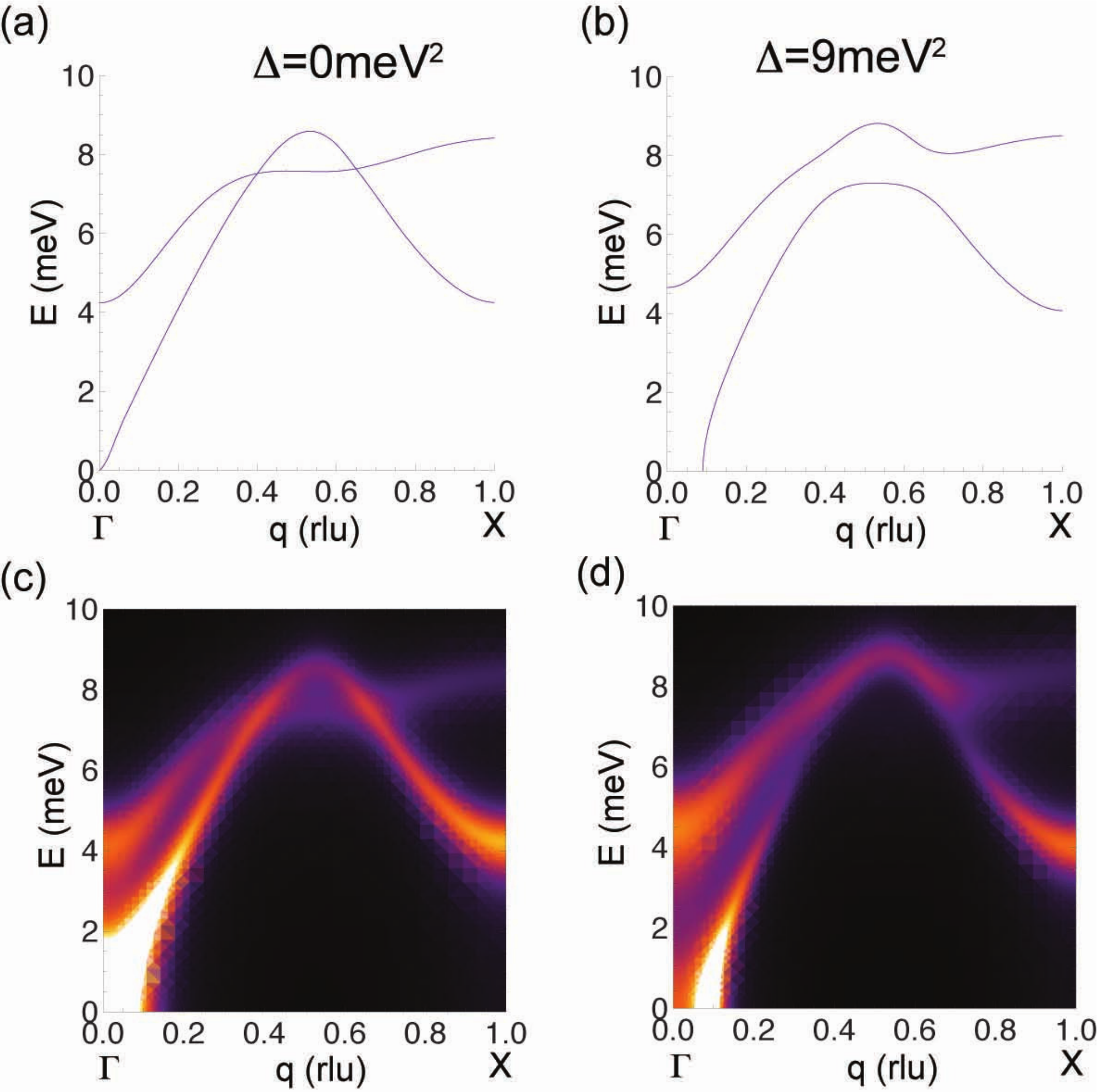}
\caption{Calculation of effect of LA-TO mode coupling on dispersion branches and scattering intensity. (a,b): dispersions, without and with coupling, respectively. (c,d): scattering intensity, without and with coupling, respectively. In both (b) and (d), the strength of the coupling is $\Delta=9\,$meV$^2$.} \label{mode-coupling}
\end{figure*}

We performed simple mode coupling calculations, to investigate the extinction of the LA branch observed in our data along [001] in the (113) zone.  We limited ourselves to a simple one-dimensional model with only LA and TO branches, and followed the model described in \cite{Hlinka}. The non-interacting harmonic dispersions were obtained from the DFT calculations described above. We investigate the effect of a simple $q$-independent coupling, $\Delta$, on the dispersions and the scattering intensity. The coupling introduces non-diagonal elements in the dynamical matrix: 
$$\mathcal{D}(q) = \left[\begin{array}{cc} {\omega_{\rm LA}^2} & {\Delta} \\ {\Delta^{*}} & {\omega_{\rm TO}^2} \end{array}\right] \quad ,$$
\noindent where $\omega_{\rm LA}(q)$ and $\omega_{\rm TO}(q)$ are the bare frequencies of the LA and TO branches at wavevector $q$, obtained with DFT. We neglect non-diagonal terms in the damping matrix, 
$$\Gamma(q) = \left[\begin{array}{cc} {\Gamma_{\rm LA}} & {0} \\ {0} & {\Gamma_{\rm TO}} \end{array}\right]  \quad ,$$

\noindent where $\Gamma_{\rm LA}$ and $\Gamma_{\rm TO}$ are the linewidths of LA and TO modes, which we take equal to $1\,$meV and $2\,$meV, respectively. At high temperature $T$, the scattering intensity is calculated as $I(q,\omega) \propto kT / \omega \;  f(q)^{*} \; {\rm Im} \, G(q,\omega) \; f(q)$, with $G(q,\omega)=(\mathcal{D}(q) - i \omega \Gamma(q) - \omega^2 \,E_2)^{-1}$, with $E_2$ a $2 \times 2$ unit matrix, and $f(q)$ a $2$-vector defining the relative strengths of LA and TO branches, which we keep equal to $[1,1]$ for simplicity. We show the results for $\Delta=0\,$meV$^2$ and $\Delta=9\,$meV$^2$ in Fig.~\ref{mode-coupling}.

The interaction produces a repulsion between the branches, and an ``avoided crossing'' where the bare phonon branches would normally cross. The interaction also produces the weakening in intensity of the LA branch near the avoided crossing, as observed experimentally. We note that our simplistic model with a constant $\Delta(q)$ would tend to produce unstable LA phonons near $\Gamma$, but we cannot entirely trust such a simple calculation. In particular, the real interaction, besides being three-dimensional, could have a non-trivial $q$ dependence.

\bibliographystyle{is-unsrt}
\bibliography{Delaire_PbTe_arxiv}{}